\documentclass[conference]{IEEEtran}
\IEEEoverridecommandlockouts
\usepackage{cite}
\usepackage{amsmath,amssymb,amsfonts}
\usepackage{graphicx}
\usepackage{textcomp}
\usepackage{xcolor}
\usepackage{hyperref}
\usepackage{footnote}
\usepackage{algpseudocode}
\usepackage{algorithm}
\usepackage{subfig}

\def\BibTeX{{\rm B\kern-.05em{\sc i\kern-.025em b}\kern-.08em
    T\kern-.1667em\lower.7ex\hbox{E}\kern-.125emX}}
\begin{document}

\title{An Improved Algorithm to Identify More Arbitrage Opportunities on Decentralized Exchanges
}

\author{\IEEEauthorblockN{Yu Zhang}
\IEEEauthorblockA{\textit{BDLT, IfI Department}, \\ \textit{University of Zurich}\\
Zurich, Switzerland 
\\Email: zhangyu@ifi.uzh.ch}
\and
\IEEEauthorblockN{Tao Yan}
\IEEEauthorblockA{\textit{BDLT, IfI Department}, \\ \textit{University of Zurich}\\
Zurich, Switzerland}
\and
\IEEEauthorblockN{Jianhong Lin}
\IEEEauthorblockA{\textit{BDLT, IfI Department}, \\ \textit{University of Zurich}\\
Zurich, Switzerland}
\and
\IEEEauthorblockN{Benjamin Kraner}
\IEEEauthorblockA{\textit{BDLT, IfI Department}, \\ \textit{University of Zurich}\\
Zurich, Switzerland}
\and
\IEEEauthorblockN{Claudio J. Tessone}
\IEEEauthorblockA{\textit{BDLT, IfI Department}, \\ \textit{University of Zurich}\\
Zurich, Switzerland}
\and
}

\maketitle

\begin{abstract}
In decentralized exchanges (DEXs), the arbitrage paths exist abundantly in the form of both arbitrage loops (e.g. the arbitrage path starts from token $A$ and back to token $A$ again in the end, $A\rightarrow B \rightarrow,..., \rightarrow A$) and non-loops (e.g. the arbitrage path starts from token$ A$ and stops at a different token $N$, $A\rightarrow B \rightarrow,..., \rightarrow N$). 
The Moore-Bellman-Ford algorithm, often coupled with the ``walk to the root" technique, is commonly employed for detecting arbitrage loops in the token graph of decentralized exchanges (DEXs) such as Uniswap.
However, a limitation of this algorithm is its ability to recognize only a limited number of arbitrage loops in each run. Additionally, it cannot specify the starting token of the detected arbitrage loops, further constraining its effectiveness in certain scenarios.
Another limitation of this algorithm is its incapacity to detect non-loop arbitrage paths between any specified pairs of tokens.
In this paper, we develop a new method to solve these problems by combining the line graph and a modified Moore-Bellman-Ford algorithm (MMBF). This method can help to find more arbitrage loops by detecting at least one arbitrage loop starting from any specified tokens in the DEXs and can detect the non-loop arbitrage paths between any pair of tokens.
Then, we applied our algorithm to Uniswap V2 and found more arbitrage loops and non-loops indeed compared with applying the Moore-Bellman-Ford (MBF) combined algorithm. The found arbitrage profit by our method in some arbitrage paths can be even as high as one million dollars, far larger than that found by the MBF combined algorithm.
Finally, we statistically compare the distribution of arbitrage path lengths and the arbitrage profit detected by both our method and the MBF combined algorithm, and depict how potential arbitrage opportunities change with time by our method.
\end{abstract}


\section{Introduction}
Decentralized finance (DeFi) has emerged as a prominent field within the blockchain ecosystem, leveraging blockchain technology to create financial applications that are open, transparent, and accessible to anyone. A crucial field of DeFi is the decentralized exchange (DEX) which allows peer-to-peer trading of tokens without the necessity of intermediaries \cite{Jani2022AnEmp}. 
 
Just like in the foreign exchange market (Forex), the prices of the same token in DEXs may differ significantly across different liquidity pools, which creates many arbitrage opportunities\cite{Jani2022AnEmp,yewa2023cyclic}. A liquidity pool is a smart contract that facilitates the exchange of two different tokens.

Among various DEXs, Uniswap is the largest DEX in terms of Total Value Locked (TVL) \footnote{Currently, it has 3.62 billion USD in TVL according to Defillama (\url{https://defillama.com/protocol/uniswap-V2)} (accessed Nov. 15, 2023)}.
Uniswap adopts the constant product market maker (CPMM) mechanism \cite{10.1145/3570639} to determine the exchange rate between a pair of tokens in a liquidity pool. Under the CPMM model, the price of each token equals the ratio of the two tokens' reserves in the liquidity pool.
The Uniswap exchanges can be mapped to a directed weighted token exchange graph $G(V, E, P)$, where nodes $(V)$ represent tokens, an edge $(E)$ denotes a liquidity pool containing the pair of tokens at the two ends of the edge, and the edge weight $(P)$ denotes the exchange rate between the two tokens in the liquidity pool. The exchange rate is determined by the reserve of each token in the liquidity pool. 

Extensive arbitrage research in DEXs has shown that Uniswap provides abundant arbitrage opportunities due to tokens' price discrepancies in different liquidity pools \cite{Jani2022AnEmp,yewa2023cyclic}.

Combined with the ``walk to the root" algorithm, the Moore-Bellman-Ford (MBF) algorithm was commonly used in detecting loop arbitrage opportunities in DEXs by recognizing negative loops\footnote{We will use the `MBF combined algorithm' simply to refer to the MBF algorithm combined with the ``walk to the root" algorithm in the later description.}. A negative loop is a loop in which the summation of all edges' weights is negative. For the arbitrage problem, an edge usually corresponds to a liquidity pool containing tokens at the two ends of the edge, and the weight of a specific edge is the log of the exchange rate of the pair of tokens in the liquidity pool. However, the drawback of the MBF combined algorithm method is that only several arbitrage loops can be recognized after each run and it can not be used for detecting non-loop arbitrage opportunities if the negative loops exist.

Against this background, this work focuses on developing a new method by combining the line graph of the token graph $G(V, E, P)$ and a modified Moore-Bellman-Ford algorithm (MMBF) to identify more arbitrage opportunities on Uniswap V2\footnote{Uniswap has 3 versions, this work primarily focuses on Uniswap V2 which has a TVL of 1.33 billion USD on November 15, 2023. Despite the launch of Uniswap V3, trading activities on Uniswap V2 are still active and the number of new pairs created is even more than before as shown in Fig. \ref{fig:number_of_pairs}.}, including not only more arbitrage loops but also non-loops between any pair of tokens.

The main contributions of this paper are summarized as follows:
\begin{enumerate}
  \item We develop an arbitrage detection algorithm that can not only find more profitable arbitrage loops than the MBF  combined algorithm but also detect arbitrage non-loop between any pair of tokens on Uniswap.
  \item We present an empirical analysis of the identified arbitrage opportunities by our algorithm and by the MBF combined algorithm.
  \item We calculate the arbitrage profit that existed in the token graph at different times by our algorithm and provide insights into the efficiency of Uniswap with time. 
\end{enumerate}
The remainder of this paper is organized as follows. Section 2 provides an overview of related work in the field of arbitrage detection on DEXs. Section 3 describes the data sources and data processing. Section 4 presents the details of our method for identifying arbitrage opportunities. Section 5 compares the arbitrage opportunities found by applying both our method and the MBF combined algorithm, including the identified arbitrage opportunities and profits. Finally, Section 6 concludes the paper with a summary of the findings and a discussion of their implications.

\section{Related Work} 

After the appearance of the DEXs for tokens, like Uniswap V2, though still in its early stage, some scientific papers have focused on analyzing the arbitrage opportunities of DEX. Wang et al.\cite{wang2022cyclic} analyze the potential cyclic arbitrage opportunities and explored arbitrage profit by traversing all triangles containing Ether (ETH, the native token of Ethereum), included on Uniswap V2. Robert et al.\cite{mclaughlin2023large} recognized the arbitrage transactions in the historic trade event log and applied Johnson’s cycle-detection algorithm to look for potential arbitrage opportunities. 
Zhou et al.\cite{zhou2021just} applied the MBF combined algorithm to recognize arbitrage loops. Danos et al.\cite{danos2021global} took the arbitrage problem as a convexity problem and applied the optimization operation to find arbitrage paths from a theoretical perspective. Berg et al. \cite{berg2022empirical} applied the method in \cite{danos2021global} on Uniswap V2 to research the efficiency of DEX by recognizing profitable arbitrage opportunities.

In the arbitrage detection field, the Moore-Bellman-Ford (or Bellman-Ford-Moore) algorithm combined with the negative detection algorithm ``walk to the root" is a vital method. Originally, MBF alone was used to find the shortest path from one specific node to all other nodes in a graph.
In case any negative loop exists, MBF fails to find the shortest path among nodes. A negative loop is characterized by the property that the cumulative weight of all edges within the loop is negative. When it comes to arbitrage in DEXs, a negative loop denotes an arbitrage loop which is similar to the cyclic arbitrage opportunities in Forex. In an arbitrage loop, if we invest a unit of some specific asset, then we can always get more units of this specific asset by trading along the arbitrage loop. Hence, much research dedicated to detecting such arbitrage loops in DEXs \cite{wang2022cyclic, mclaughlin2023large,jin2022detecting, hansson2022arbitrage, zhou2021just}. 

However, a very significant shortcoming of looking for arbitrage by applying the MBF combined with the negative detection algorithm ``walk to the root" is that it can only recognize very few arbitrage loops each time, and we can not even specify the starting token for the arbitrage loop. 
Besides arbitrage loops, another critical point is whether we can find a shorter non-loop path between a pair of nodes correctly. If a shorter non-loop between a pair of tokens is found, we can also get arbitrage profit by combining the path on DEX and the tokens' prices from the centralized exchanges (CEX). The MBF combined algorithm is not suitable directly on token graphs for looking for non-loop paths between a pair of nodes if negative loops exist in the network.


\section{Data Description}
\subsection{Data Source}
The token pairs that can be exchanged with each other are also called token liquidity pools in this paper. Their information on Uniswap V2 is obtained from the \textit{The Graph} \footnote{https://thegraph.com/hosted-service/subgraph/ianlapham/uniswap-V2-dev}, which provides detailed information for each token of the liquidity pools, such as tokens' addresses, names, symbols, reserves, and decimals. This information is then used to construct a token graph with nodes representing tokens and edge weights representing the exchange rates between two corresponding tokens. To identify potential arbitrage opportunities with time, we retrieve historical daily snapshots from \textit{The Graph}, which encompass the daily trading volume in USD and reserves of each token in the liquidity pool. The exchange rate of each pair of tokens is determined by their reserve ratio in the same liquidity pool. Furthermore, to calculate the profits gained in USD for arbitrage opportunities, we also collect the tokens' prices from CEXs. Finally, the analyzed data spans from 1st September 2020 to 31st October 2023.

\begin{figure}[!ht]
    \centering
    \includegraphics[width=0.5\textwidth]{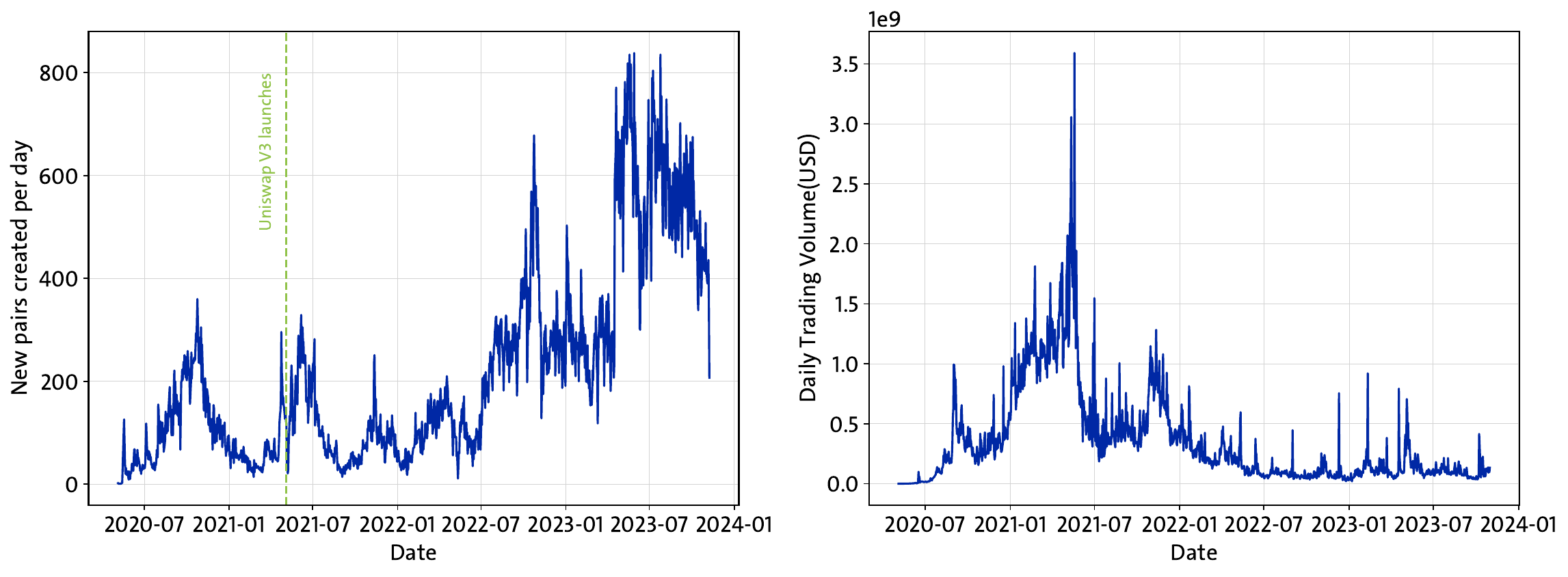}
    \caption{New pairs created and trading volume per day. The daily number of newly created pools after the launch of Uniswap V3 continued to increase and surged to 678 even on October 24th, 2022, and further spiked to 835 on July 25th, 2023. Moreover, despite the daily trading volume declining a bit following the launch of Uniswap V3, it still maintained a consistently high level and occasionally exceeded five hundred million USD.}
    \label{fig:number_of_pairs}
\end{figure}

Fig. \ref{fig:number_of_pairs} provides holistic statistics about the daily number of newly created pools and daily transaction volume in Uniswap V2.

\subsection{Data Processing}\label{data_processing}
To provide a comprehensive overview of the market on any given day, we generated the daily token liquidity pool snapshot of the whole market first. There have been more than one hundred thousand tokens as of October 31, 2023. The tokens' liquidity pool data can be used to construct the token graph. In the token graph, nodes present tokens, and an edge is linked between two tokens if there is a liquidity pool containing both of them simultaneously. In this way, all token pools can be connected as a token graph $G(V, E, P)$ with $V$ as the token set, $E$ as the token liquidity pool set, and $P$ as the exchange rate set.
However, the computation complexity is very high if all tokens are contained in our analysis, and we can not arbitrage if a token's degree is one (meaning there is only one pool that contains the specific token), we need to filter pools or tokens when building our token graph. We take the following steps to filter tokens and pools:
\begin{itemize}
    \item Calculate the first trading date and the last trading date for each pool. This is to filter active pools in constructing the token graph.

    \item On each day, we select only those pools whose Total Values Locked (TVL) are larger than twenty thousand dollars and that are still active (meaning pools whose first trading date is on or before the specific date but whose last trading date is equal or later than this specific date) to build the token graph.

    \item After the graph is constructed, we delete all tokens from the graph whose degree is one.

    \item The number of tokens in the graph is set as one hundred. If the total number of tokens in the graph is larger than one hundred, then we delete pools with the smallest TVLs. Deleting pools may trigger the deleting of other pools or tokens. So, the process of deleting pools is iterative. After one pool with a smaller TVL is deleted, we check whether the degree of the tokens that are in the deleted pool but still in the token graph is less than 2. In case some token's degree is less than 2, then we delete the corresponding token and the pool that contains the deleted token again. Because a new pool is deleted, we use the same steps above to delete tokens and pools again. 
    
    \item Run the above procedures of deleting tokens and pools from the graph until there are only one hundred tokens left in the graph. 
\end{itemize}

The above is the procedure of how we construct the token graph. After filtering, the graph has one hundred tokens and roughly the first four hundred liquidity pools with the most liquidity. 
The first largest four hundred liquidity pools in TVL take about 90\% of the total TVL and correspond to about one hundred different tokens. This is why we chose one hundred tokens and about four hundred liquidity pools in our research.

\section{Method}

Our method in arbitrage loop and non-loop detection mainly includes 4 steps. Firstly, we need to construct a token graph $G(V, E, P)$ based on the tokens' liquidity pool data on Uniswap V2. Secondly, we construct the line graph $L(G)$ according to the token graph $G(V, E, P)$. Thirdly, we introduce a $MMBF$ method and apply it on $L(G)$ to detect the arbitrage loop and non-loops. At last, for each arbitrage loop and non-loop, we use the bisection method to maximize the arbitrage profit.

\subsection{Constructing Token Graph}

$G(V,E,P)$ is a weighted directed token graph, where $v_i \in V$ is the $i^{th}$ token in the token set $V$ ($i=1,2,...N$), and $e_{ij}$ is the edge between token $v_i$ and $v_j$. $W$ denotes the edge weight set. The edge $e_{ij}$ represents a directed exchange by paying token $v_i$ to get $v_j$. It can be expressed as $e_{ij}=(v_i, v_j)$. For any pair of tokens $v_i$ and $v_j$, there exist two different edges, the edge from token $v_i$ to $v_j$ ($e_{ij}$) and the edge from token $v_j$ to $v_i$ ($e_{ji}$), so, $e_{ij}\neq e_{ji}$. Each edge has a corresponding edge weight. We use $p_{ij}$ ($\in P$), the price of token $v_i$ in the unit of token $v_j$, to denote the weight of edge $e_{ij}$. It can be calculated simply by taking a negative log to the ratio between the reserve of the token $v_i$ and $v_j$. So, $p_{ij}=-log((1-\lambda)\frac{r_j}{r_i})$, where $r_i$ and $r_j$ is the reserve of token $v_i$ and $v_j$ in the liquidity pool, respectively. $\lambda$ is the imposed tax rate by the liquidity pool. In the case of Uniswap 2 \cite{Adams2020UniswapVC}, $\lambda =0.3\%$. Based on this formula, we know that $p_{ij}  +p_{ji} = -2log(1-\lambda)$. 

We can construct the token graph $G(V, E, P)$ based on what we described here and the token filtering method in Section \ref{data_processing}.

\subsection{Constructing Line Graph for Token Graph}\label{construct_line_graph}

Given the underlying graph $G(V, E, P)$, its line graph $L(G)$ can be constructed by the following steps:
\begin{itemize}
    \item The edges ($e_{ij}$) in the original underlying graph $G(V, E, P)$ are taken as the new vertices of the line Graph $L(G)$. A new vertex in the line graph $L(G)$ will be denoted as a tuple with its two entries representing the two tokens at the two ends of each edge ($e_{ij}$) in $G(V, E, P)$. For example, in $G(V, E, P)$, the edge $e_{ij}$ is a directed edge from $v_i$ to $v_j$. In $L(G)$, the corresponding new vertex will be $(v_i, v_j)$, $i, j = 1,2,...N $ and $i\neq j$.
    
    \item Any pair of new vertices in line graph $L(G)$ are linked if the last token of a new vertex is the first token of another new vertex. For example, assuming one new vertex in $L(G)$ is $(v_i, v_j)$ and another new vertex is $(v_j,v_l)$, then we add a link from the new vertex $(v_i, v_j)$ to vertex $(v_j,v_l)$ in $L(G)$. The new edge from $(v_i, v_j)$ to $(v_j,v_l)$ in $L(G)$ can be denoted as $(v_i, v_j, v_l)$ or $(v_i, v_j)\rightarrow (v_j, v_l)$.
    In this step, we also need to set weights for all edges in $L(G)$. The weight of edge $(v_i, v_j, v_l)$ (or $(v_i, v_j)\rightarrow (v_j, v_l)$) in $L(G)$ is set as $p_{jl}$ which is the edge weight from token $v_j$ to $v_l$ in $G(V, E, P)$.
    
     \item For any pair of new vertices in $L(G)$, $(v_i, v_j)$ and $(v_k,v_l)$, if $j = k$, $i=l$ and $i \neq j$, then we cut both the two mutual links between the two vertices. This means we cut both the edge $(v_i, v_j, v_i)$ and edge $(v_j, v_i, v_j)$ in $L(G)$. By cutting these edges, we can reduce the computation complexity because the arbitrage loops between a pair of tokens, like $v_i\rightarrow v_j \rightarrow v_i$ and $v_j\rightarrow v_i \rightarrow v_j$, are non-profitable.
\end{itemize}

The number of vertices in the new line graph ($M_{L(G)}$) equals the number of edges in the underlying graph ($E_G$), namely, $M_{L(G)}=E_G$. The number of edges in the line graph ($E_{L(G)}$) equals the sum of the degree's square of each node minus two times the number of edges in graph $G$, namely $E_{L(G)} =\sum {d_i}^2-2E_G$, where $d_i$ denotes the degree of token (node) $i$ in the underlying graph $G$ and $E_G$ denotes the number of edges of the underlying graph $G$.

\subsection{Modified Moore-Bellman-Ford (MMBF) on Line Graph with Extra Node}

According to the line graph constructing rules, each token in the original underlying graph $G(V, E, P)$ is included in several different new vertices in the line graph $L(G)$. For example, assuming $e_{ij}$ and $e_{ik}$ are two edges starting from token $v_i$ in $G(V,E,P)$, both vertices $(v_i,v_j)$ and $(v_i,v_k)$ in $L(G)$ contain the token $v_i$.
For the sake of simplicity, we call these vertices ($(v_i,v_j)$ and $(v_i,v_k)$) in $L(G)$ the `neighbor link vertices' of $v_i$, and that $v_j$ and $v_k$ constitute to the `neighbor token set' of $v_i$. These two concepts will facilitate our description below.

We assume that $N_i$ ($v_{n_1}, v_{n_2},...v_{n_i} \in N_i$) is the neighbor token set of the specific token (source token) $v_i$. Based on the constructing rule of line graph $L(G)$, $v_i$ has $n_i$ neighbour link vertices in  $L(G)$, namely, $(v_i, v_{n_1})$, $(v_i, v_{n_2})$, $...$, $(v_i, v_{n_i})$.

To detect the arbitrage loop and non-loops, the MMBF can be used in two different ways in the line graph $L(G)$. One way is to calculate the shortest path between the source token's ($v_i$) neighbor link vertices ($N_i$) and all other vertices in $L(G)$. This method is very time-consuming because the MMBF will be applied once for each neighbor link vertex. 

We now present the second efficient way that we will apply in the paper. The second way simplifies the computation by adding an extra node to the line graph $L(G)$ first and then applying the MMBF only once. The addition of an extra node is as follows:

 \begin{itemize}
     \item Construct the line graph $L(G)$ by steps as shown in section \ref{construct_line_graph}.
     \item We assume that $v_0$ is the source token that we will consider and $(v_0, v_{n_1})$, $(v_0, v_{n_2})$, $...$, $(v_0, v_{n_k})$ are its neighbour link vertices in $L(G)$. Now, we add an extra node called $(O,v_0)$ to $L(G)$ and link this vertex to all the neighbor link vertices of $v_0$, namely, $(O,v_0)\rightarrow(v_0, v_{n_2})$, $...$, $(O,v_0)\rightarrow(v_0, v_{n_k})$. $(O,v_0)$ is also called the source vertex in $L(G)$, and $O$ is just a sign that is not important and can be any other sign.
     \item The weight of each extra added edge in $L(G)$ is set as the weight of the edge from the source token ($v_0$) to all tokens in the neighbor token set of $v_0$ in $G(V, E, P)$, respectively. For example, the weight of the edge $(O,v_0)\rightarrow(v_0, v_{n_k})$ is $p_{v_0v_{n_k}}$.
 \end{itemize}

Then we use the modified Moore-Bellman-Ford algorithm to calculate the shorter paths from the source vertex $(O,v_0)$ to all other vertices in the line graph $L(G)$. The MMBF algorithm is described in Algorithm \ref{alg:cap1}.
\begin{algorithm}[h]
\caption{Modified Moore-Bellman-Ford algorithm (MMBF)}   \label{alg:cap1}
\small{
    \begin{algorithmic}
    \State $M \gets len(G)$: the number of tokens in $G$
    \State Line graph: $L(G)$ 
    \State Source vertex: $(O,v_0)$
    \State Edge weight of $L(G)$: $W$
    \State Edge weight of $G$: $P$
    \State Define $Dis$ and $Path$ as two dictionaries to store distances and arbitrage paths from the source vertex $(0,v_0)$ to all other vertices in $L(G)$.
\vspace{0.1cm}
    \For{$m =1,2,..., M$}
        \For{Each edge $(v_i, v_j)\rightarrow (v_j, v_l)  $ in $L(G)$}
        
            \If{$Dis_{(v_i, v_j)}+ W_{(v_i, v_j)\rightarrow (v_j, v_l)}< Dis_{(v_j, v_l)}$ \\and ($v_l \notin Path_{(v_i, v_j)}$ or $l = 0$)} 
            \\
            \State $Path_{(v_j, v_l)}=Path_{(v_i, v_j)}+(v_j, v_l)$,\\
            $Dis_{(v_j, v_l)}=Dis_{(v_i, v_j)}+ W_{(v_i, v_j)\rightarrow (v_j, v_l)}$
            \EndIf
            
        \EndFor
    \EndFor
    \vspace{0.1cm}
    \State Define $D_{token}$ and $P_{token}$ as two dictionaries to store distances and arbitrage paths from the source token $v_0$ to all other tokens in $G$.
    \For{Each key-value pair $k_d, v_d $ in $Dis$}
        \State $t=k_d[-1]$
        \If{$v_d< D_{token}[t]$}
            \State $D_{token}[t]=v_d$
            \State $P_{token}[t]=Path_{k_d}$
        \EndIf
    \EndFor
    
    \Return $D_{token}$, $P_{token}$
    \end{algorithmic}
    }
\end{algorithm}

By applying MMBF on $L(G)$ with an extra node, we can detect the arbitrage non-loops between any pair of tokens. At the same time, we can also get an arbitrage loop starting from the source token and back to the source token. For example, if the source token in is $v_0$, and other tokens are $v_1,v_2,...,v_n$, we can get all arbitrage non-loops between $v_0$ and all other tokens $v_1,v_2,...,v_n$. At the same time, we can also get the arbitrage loop starting from $v_0$ and backing to $v_0$ again. If we take different tokens as the source token and repeat all the processes above, we can get the arbitrage non-loop between any pair of tokens and arbitrage loops from each token in the token graph $G(V,E,P)$ finally. For a token graph with $N$ tokens, the method can get $N$ arbitrage loops and $N^2$ non-loops.

\subsection{Adjusting Input to Maximise the Arbitrage Profit}

After we find a profitable loop or non-loop, the next question is how much we should invest to maximize our arbitrage profit. This question is closely correlated with the tokens' exchange rates and liquidity depth in each liquidity pool. 

The constant Automate Market Marker (AMM) equation in the liquidity pool of Uniswap version 2 is:
\begin{equation}
        [x+(1-\lambda) \Delta x](y-\Delta y) = xy=k,
        \label{eq1}
\end{equation}
where $x$ and $y$ denotes the number of token $x$ and $y$ in the liquidity pool, respectively. $\lambda$ denotes the transaction tax rate. 

From the above equation, we know that $\Delta y$ is a concave function that increases monotonically with $\Delta x$. As well known, if a function $F(\cdot)$ is concave, then $F(G(\cdot))$ is also concave if function $G(\cdot)$ is concave. When the path includes more tokens, the quantity of inputs for the starting token continues to exhibit a concave and monotonically increasing pattern concerning the number of target tokens being extracted. 
So, from the economic point of view, when the marginal output of the target token for each unit of the starting token is equal to the reserve ratio of the target token ($R_y$) to the starting token ($R_x$) in case that a liquidity pool containing both $x$ and $y$ exists, or equal to the market price of the token $x$ to that of token $y$ from the centralized market, the arbitrage profit is maximal. In the case of an arbitrage loop, the result is still applicable. When it is in terms of loops, the starting token and the target token are the same, so, they have the same price and we get the maximal profit when $\frac{d (\Delta y)}{d(\Delta x)}=1$.
\vspace{-0.5cm}
\begin{figure}[!ht]
    \centering
    \includegraphics[width=0.35\textwidth]{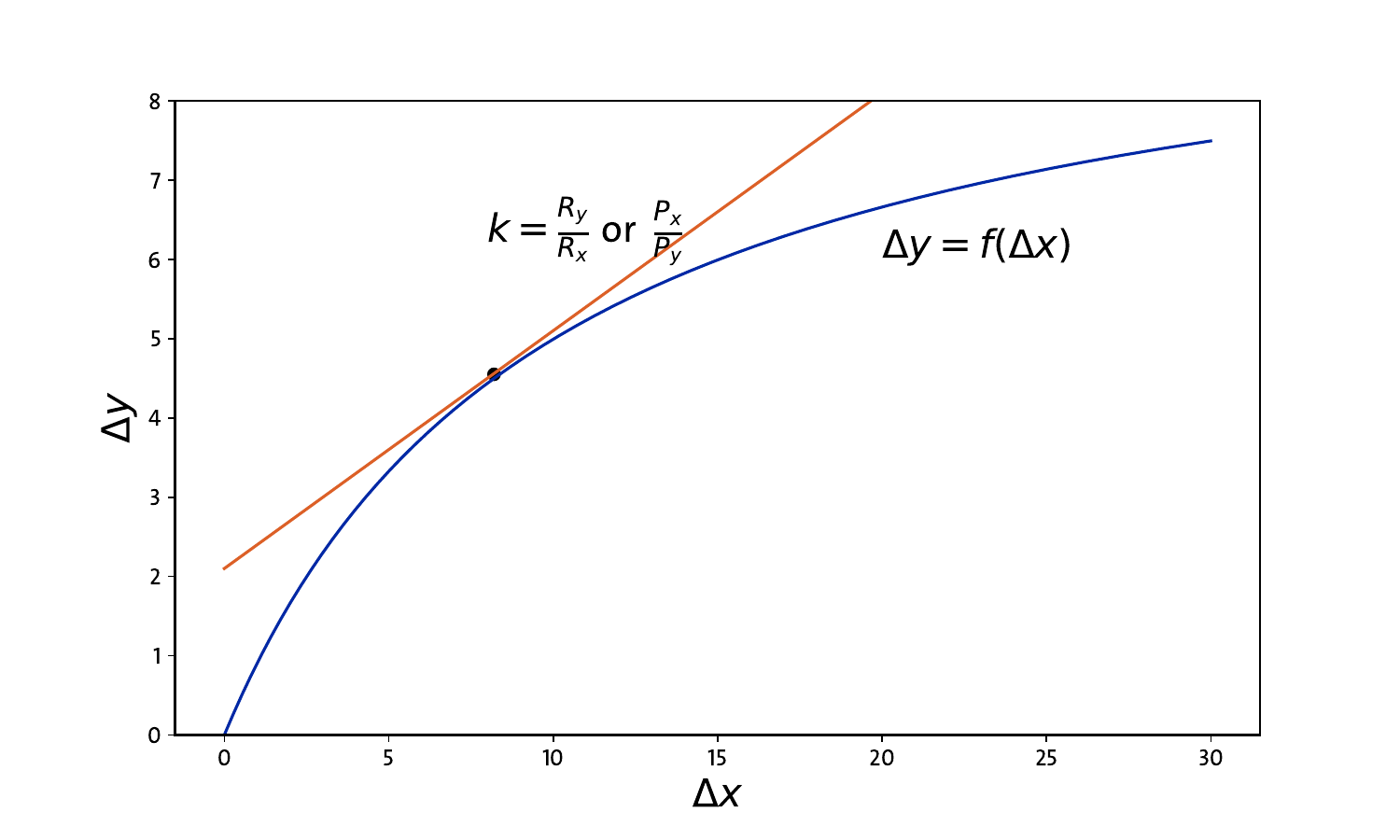}
    \caption{The profit is maximal when the marginal output of the target token ($y$) we get for a unit of starting token ($x$) is equal to the ratio of the target token reserve ($R_y$) to the starting token reserve ($R_x$), or equal to the two tokens' market price ratio ($\frac{P_x}{P_y}$) with the prices from CEXs.}
    \label{fig:enter-label}
\end{figure}
\setlength{\textfloatsep}{5pt}
Because of the concave and monotonically increasing function, it is very easy and efficient to use the bisection method to calculate the optimal input of the starting token to get maximal arbitrage profit.

\section{Statistics of the Potential Arbitrage on Uniswap V2 by Our Method Comparing to the MBF Combined Algorithm}
In this section, we summarise several important statistical findings by applying our methods to historical data on Uniswap V2 and then compare the results to those by applying the MBF combined algorithm. 

\begin{figure}[!ht]
\centering
\subfloat[MMBF]{\includegraphics[width=0.4\textwidth]{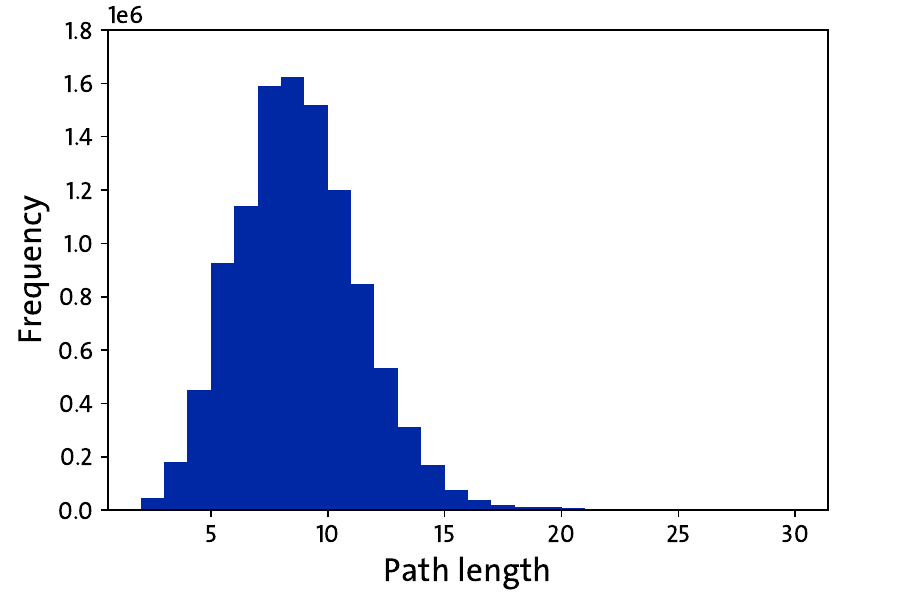}\label{fig_first_case0}}
\hfil
\subfloat[MBF]{\includegraphics[width = 0.4\textwidth]{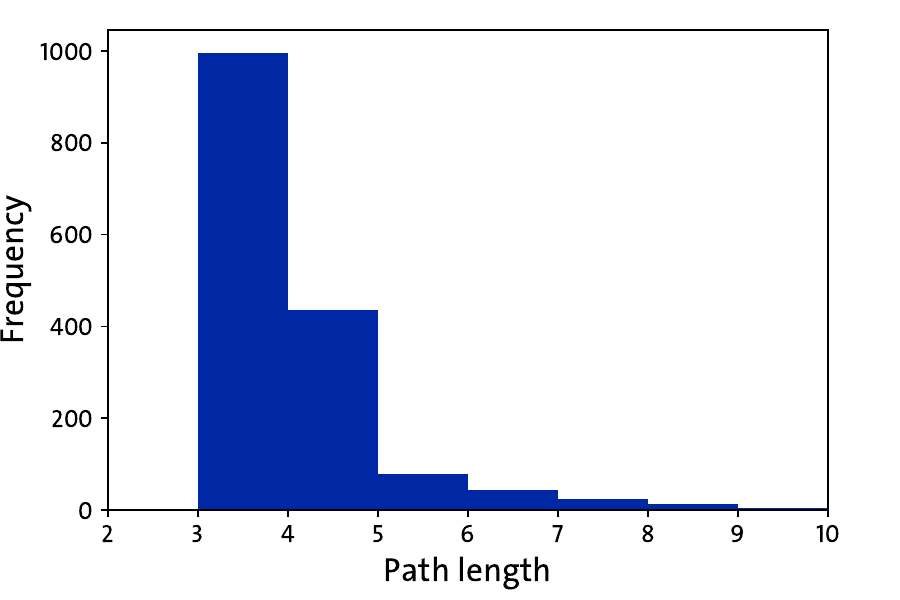}\label{fig_second_case1}}
\caption{Distribution of the length of arbitrage paths. Panel \ref{fig_first_case0} and \ref{fig_second_case1} depicts the distribution of arbitrage path length by our method and the MBF combined algorithm, respectively. }
\label{path_distri}
\end{figure}

First, we depicted the distribution of the arbitrage path length of arbitrage paths found by both our method and the MBF combined algorithm, which is shown in Fig. \ref{path_distri}. The number of arbitrage loops and non-loops by our method is much larger than that by the MBF combined algorithm, which denotes that our method can find more arbitrage opportunities than the MBF combined algorithm. The distribution of arbitrage paths by our method is nearly symmetric, and we find that the length of most arbitrage paths ranges from 7 to 11. The profitable paths whose length ranges from 3 to 6 and from 12 to 15 also take a very large percentage. The distribution of arbitrage loop length by the MBF combined algorithm is a lateral distribution in that most arbitrage loops have lengths of 3 and 4. 

\begin{figure}[!ht]
\centering
\subfloat[MMBF]{\includegraphics[width=0.4\textwidth]{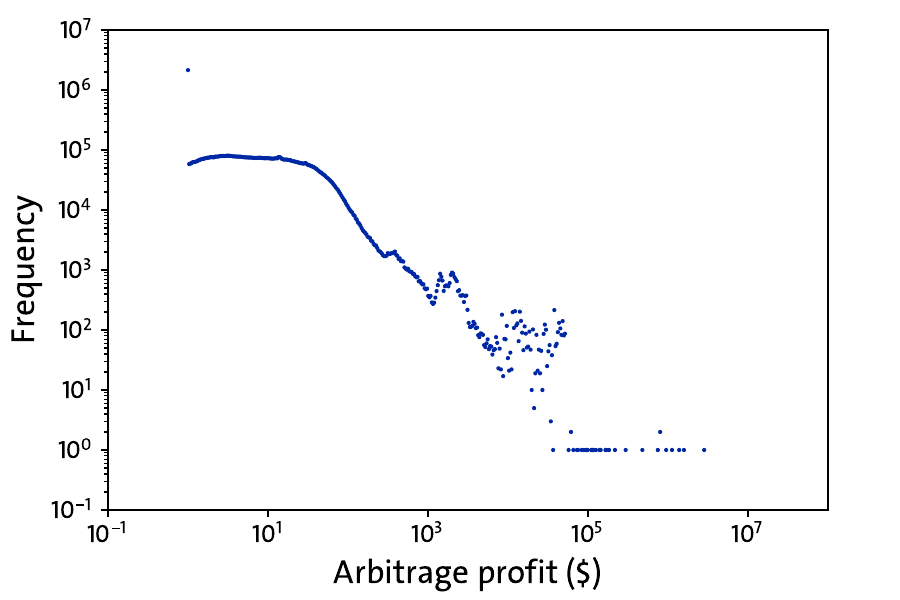}\label{fig_first_case2}}
\hfil
\subfloat[MBF]{\includegraphics[width = 0.4\textwidth]{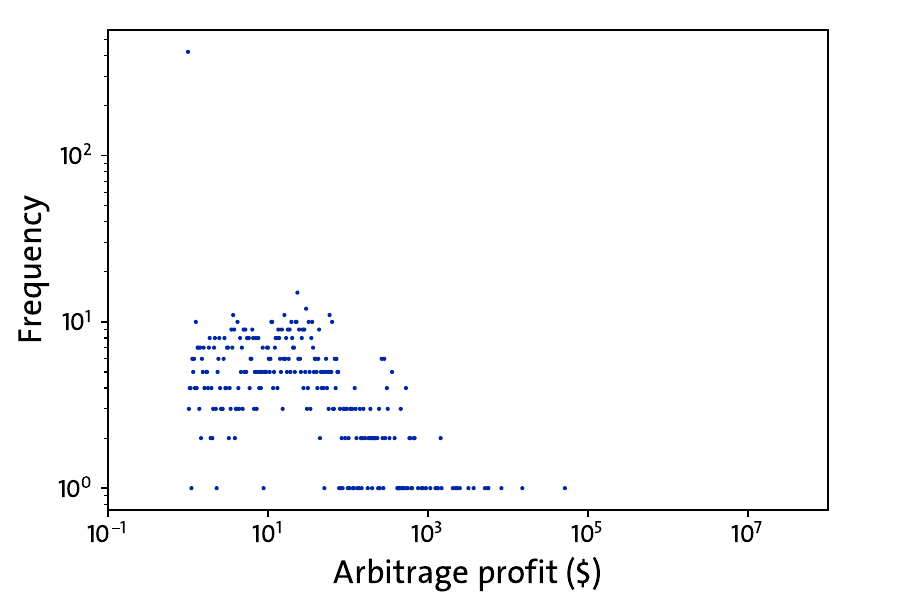}\label{fig_second_case3}}
\caption{Distribution of potential arbitrage profit in the token graph. Panel \ref{fig_first_case2} and \ref{fig_second_case3} depicts the distribution of potential arbitrage profit by applying our method and the MBF combined algorithm, respectively.}
\label{profit_distri}
\end{figure}

\begin{figure}[!ht]
    \centering
    \includegraphics[width=0.4\textwidth]{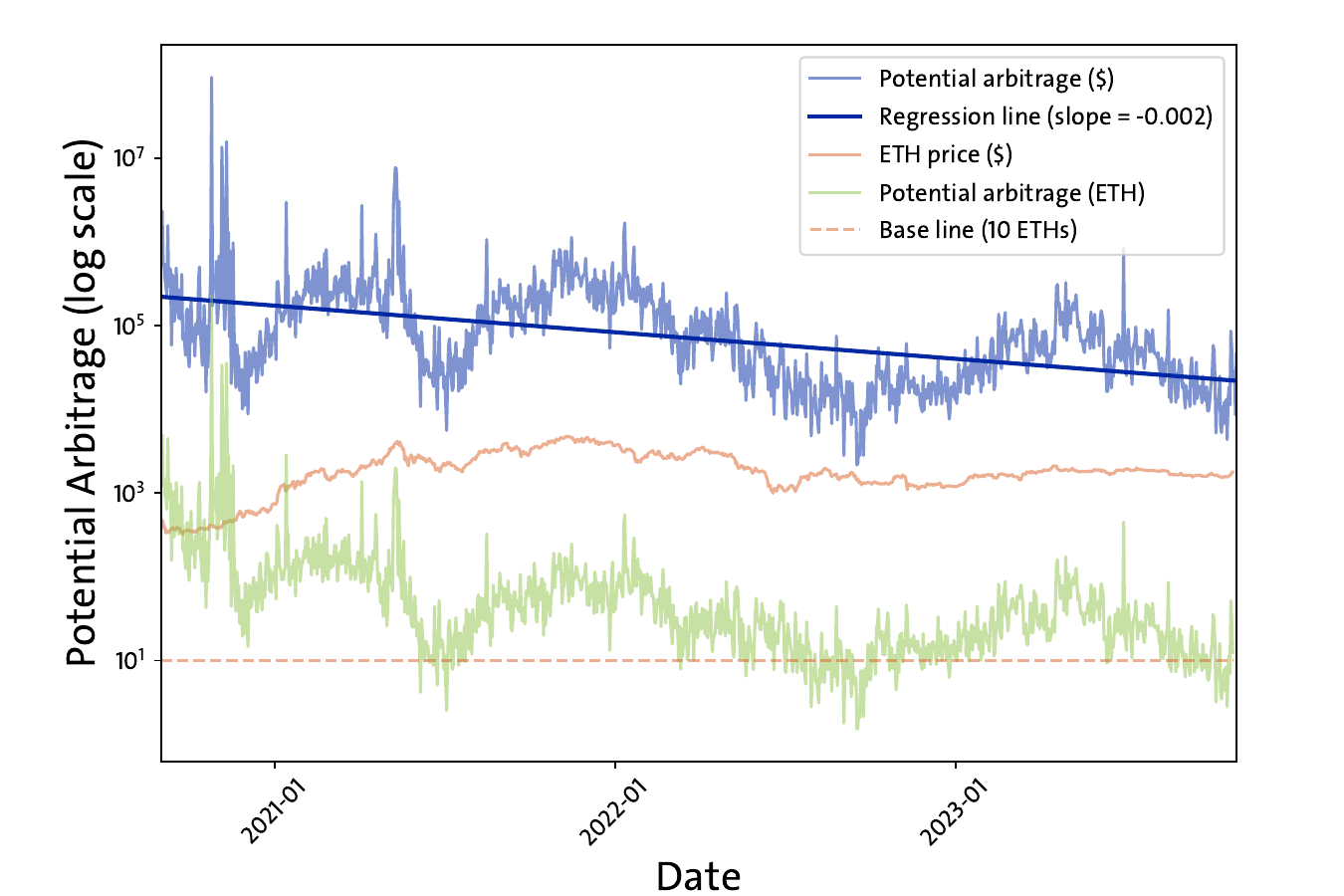}
    \caption{Potential profit from arbitrage by our method in Uniswap V2 with time.}
    \label{arb_profit_time}
\end{figure}

\setlength{\textfloatsep}{5pt}
Second, we depicted the distribution of potential profit in all arbitrage paths by applying our method and the MBF combined algorithm. Still, more arbitrage profits are found by our method than by the MBF combined algorithm, as shown in Fig. \ref{profit_distri}. The largest profit by our method can be a million dollars, compared to about one hundred thousand by the MBF combined algorithm. 23,868 arbitrage paths with more than a thousand dollars are found by our method, the number is only 19 by the MBF combined algorithm. We can find that a larger number of arbitrage possibilities are less than ten dollars. If we focus on the arbitrage opportunities whose profits are larger than one hundred dollars, we find that they follow almost a power-law distribution by our method. The arbitrage profit in most paths is hundreds of dollars. By checking the profit distribution, we find several paths in which the arbitrage profit can be up to almost one million dollars. We need to be aware that this potential arbitrage profit in this paper is only calculated in snapshots, which means that arbitrageurs can always find profitable paths or loops at any time.
\setlength{\textfloatsep}{0pt}
Finally, we plot how the total detected potential arbitrage profit that exists in Uniswap V2 evolves with time by our method, which is shown in Fig. \ref{arb_profit_time}. Because the total arbitrage profit by the MBF combined algorithm is small, we do not plot it here. From Fig. \ref{arb_profit_time}, we can see that the total profitable arbitrage fluctuates. At the end of 2020, much potential profit existed in the Uniswap V2 network and it can be as high as ten million dollars on some days. Another finding is that the total profitable arbitrage seems to be cyclical. There are several large cycles (troughs and peaks). By making a rough and simple line regression, we find that the slope coefficient is about $-0.002$, which denotes that the total profitable arbitrage has a trend to decline with time. We take this as an indication that the market mechanisms of DeFi, namely the AMM studied, Uniswap V2, are being arbitraged more efficiently, and thus Uniswap V2 can provide better price information to market participants. Further studies could involve comparing the volume of TVL locked-in tokens with arbitrage opportunities and relating the depth of these markets to the possible arbitrage.  

\section{Result and Discussion}
In this study, we integrate the line graph approach with a modified MBF algorithm to propose a novel method for identifying arbitrage opportunities in token networks. Compared with the MBF combined algorithm, an advantage of our method is that we can not only find valid and shorter paths for any pair of tokens but also identify loops originating from any source node. Another advantage is that our method can recognize more arbitrage opportunities than the MBF combined algorithm. By systematically applying our algorithm to each token in the network as a source, we can enumerate potential loops starting from each token and non-loops for every token pair, along with their respective maximum profits through optimal input token allocation.

We applied our algorithm to historical data from Uniswap V2 (1st September 2020 to 31st October 2023) and found that a large number of potential arbitrage opportunities exist, which shows that our algorithm is effective in finding arbitrage opportunities. For example, more than a million dollars of arbitrage existed in the graph network at the end of 2020. In May of 2023, the potential arbitrage profit is still more than one hundred thousand dollars. Furthermore, we show that the total dollar value of these arbitrage opportunities decreased with time, which prompts further research exploration.

We also tested the running time of our method in detecting the arbitrage paths in the token graph with one hundred tokens and about four hundred liquidity pools, and the running time is about 8-10 seconds, which means that it can be used in practical arbitrage detection. If the code is optimized again, we believe that our method can be faster. In a real case, we don't need to construct a token graph all the time, we just need to update the tokens' balance based on the latest transaction data, and then run the MMBF again, which means that this method can be faster in the real case. 

However, our approach is not without limitations and needs to improve. While our algorithm can find the valid arbitrage path between a pair of tokens compared with the MBF combined algorithm even in cases where many negative loops exist, this arbitrage path may not be the most profitable. This topic is very challenging in theory and will be the subject of our future research. Another limitation of this paper is that the gas fee is not considered. The gas fee is a significant factor that deserves to be considered in on-chain transactions. For simplicity, we can take it as a fixed cost, like fifty dollars for each transaction, or increase the tax rate $\lambda$ in equation \ref{eq1}, both of which will not change the analysis much in the paper. 

\bibliographystyle{IEEEtran}
\bibliography{reference.bib}

\nocite{*}

\end{document}